\documentclass[11pt,a4paper]{article}

\usepackage[margin=2.5cm]{geometry}
\usepackage{graphicx}
\usepackage{booktabs}
\usepackage{hyperref}
\usepackage{listings}
\usepackage{xcolor}
\usepackage{amsmath}
\usepackage{caption}
\usepackage{subcaption}
\usepackage{float}
\usepackage{enumitem}
\usepackage{titlesec}
\usepackage{parskip}
\usepackage{microtype}
\usepackage{fancyhdr}
\usepackage{abstract}
\usepackage{amssymb}

\hypersetup{colorlinks=true, linkcolor=blue!60!black, urlcolor=blue!60!black, citecolor=blue!60!black}

\definecolor{codebg}{rgb}{0.96,0.96,0.96}
\definecolor{codeframe}{rgb}{0.85,0.85,0.85}
\definecolor{kwcolor}{rgb}{0.0,0.3,0.6}
\definecolor{strcolor}{rgb}{0.2,0.5,0.2}
\definecolor{cmtcolor}{rgb}{0.5,0.5,0.5}

\title{\textbf{Fine-Tuning and Serving Gemma 4 31B on Google Cloud TPU}\\A Technical Comparison with GPU Baselines}

\author{
  Jatin Kishnani \quad Mayank Goel \quad Amit Singh \quad Pulkit Agrawal \\[5pt]
  Sairanjan Mishra 
}

\begin{document}

\maketitle
\thispagestyle{fancy}

\begin{abstract}
We present the first end-to-end demonstration of fine-tuning and serving Google's Gemma 4 31B model on TPU hardware, providing an empirical comparison of TPU and GPU platforms for large language model adaptation. Using LoRA on a Google TPU v5p-8 for training and TPU v6e-8 (Trillium) for inference, we document the full set of code-level adaptations required to port a GPU-native training recipe - built on PyTorch, HuggingFace TRL, and FSDP - to the JAX + Tunix/Qwix stack. These adaptations span mesh configuration, LoRA module naming conventions, sharding annotation corrections, gradient checkpoint, data pipeline restructuring, and a custom Orbax-to-safetensor checkpoint merging procedure. For inference, we detail the vLLM-TPU Docker setup necessary to serve Gemma 4 on v6e-8 and characterize the resulting latency and throughput profile. Compared with a similar-costing 2×H100 GPU baseline under identical hyperparameters, TPU training completes 1.61× faster at 2.12× lower cost. For inference, we cover the vLLM-TPU Docker setup required to serve Gemma 4 on v6e-8 and explain the observed latency and throughput characteristics across a QPS sweep spanning 512 to 16k input tokens. Across both workloads we compare performance and cost against a 2$\times$H100 GPU baseline running identical hyperparameters. The TPU completes training 1.61$\times$ faster at 2.12$\times$ lower cost. For inference, TPU v6e-8 matches GPU at short context ($\leq$2048 tokens) and decisively outperforms at long context: 66\% higher throughput and 23.6$\times$ faster TTFT at 4096-token inputs (61\,ms vs 1,443\,ms at QPS=4). Our work removes a critical gap in the open tooling ecosystem and provides practitioners with a recipe for Gemma 4 Dense 31B deployment on the TPU infrastructure.

\end{abstract}

\tableofcontents
\newpage

%% ─────────────────────────────────────────────────────────────────────────────
\section{Introduction}
%% ─────────────────────────────────────────────────────────────────────────────

Large language model fine-tuning is increasingly bottlenecked not by data availability but by compute cost and iteration speed. Google Cloud TPUs, while natively supported for training, require non-trivial porting effort compared to the mature GPU ecosystem (PyTorch, HuggingFace, FSDP). This report documents our end-to-end experience running Gemma 4 31B\footnote{Base model: \url{https://huggingface.co/google/gemma-4-31B}} LoRA supervised fine-tuning on TPU v5p-8 and vLLM inference on TPU v6e-8 (Trillium), and compares both to an existing GPU baseline on 2$\times$H100 80GB (GCP \texttt{a3-highgpu-2g}).

The task is Verilog code generation: we fine-tune on the CodeV-R1 dataset~\cite{codev} (10K samples of natural-language-to-Verilog pairs) and evaluate on NVlabs/verilog-eval~\cite{verilogeval} (156 spec-to-RTL problems, pass@1/pass@5 with iverilog simulation).

The paper is structured as a walkthrough of the engineering changes required to make TPU work, followed by quantitative results and cost analysis.

%% ─────────────────────────────────────────────────────────────────────────────
\section{Hardware and Infrastructure}
%% ─────────────────────────────────────────────────────────────────────────────

\subsection{Training Hardware}

\begin{table}[H]
\centering
\caption{Training hardware configurations.}
\begin{tabular}{lll}
\toprule
\textbf{Spec} & \textbf{TPU v5p-8} & \textbf{GPU a3-highgpu-2g} \\
\midrule
Accelerator & 4$\times$ TPU v5p chips & 2$\times$ NVIDIA H100 80GB HBM3 \\
HBM per chip & 102.8 GB & 80 GB \\
Total HBM & 411.2 GB & 160 GB \\
Host RAM & \textasciitilde355 GB & \textasciitilde468 GB \\
Interconnect & ICI (inter-chip) & NVLink \\
GCP on-demand rate & \$16.80/hr & \$22.12/hr \\
\bottomrule
\end{tabular}
\end{table}

\subsection{Inference Hardware}

\begin{table}[H]
\centering
\caption{Inference hardware configurations.}
\begin{tabular}{lll}
\toprule
\textbf{Spec} & \textbf{TPU v6e-8 (Trillium)} & \textbf{GPU a3-highgpu-2g} \\
\midrule
Accelerator & 8$\times$ TPU v6e chips & 2$\times$ NVIDIA H100 80GB HBM3 \\
HBM per chip & 31.25 GB & 80 GB \\
Total HBM & 250 GB & 160 GB \\
Tensor parallelism & tp=8 & tp=2 \\
GCP on-demand rate & \$21.52/hr & \$22.12/hr \\
\bottomrule
\end{tabular}
\end{table}

\subsection{TPU VM Environment}

Both TPU generations (v5p and v6e) ship with Python 3.10, but the Tunix framework requires Python 3.11+. The first setup step is therefore installing Python 3.11 via the \texttt{deadsnakes} PPA before anything else. The JAX TPU wheel is installed from a separate Google-hosted index:

\begin{lstlisting}[language=bash]
pip install "jax[tpu]" -f https://storage.googleapis.com/jax-releases/libtpu_releases.html
\end{lstlisting}

A critical operational note: the TPU VM's boot disk (97 GB) cannot hold the 62.5 GB Gemma 4 31B model download alongside the OS. Tunix stages the model in \texttt{/tmp/models} before loading into HBM. On the v5p-8, the host has 355 GB of RAM-backed \texttt{/dev/shm} (tmpfs), which we redirect the staging to via \texttt{TMPDIR=/dev/shm} at launch time. Without this, the model download fills the boot disk and crashes mid-load.

For the v6e-8 (Trillium) inference instance, we found that only the \texttt{v2-alpha-tpuv6e} runtime works correctly---the generic \texttt{tpu-ubuntu2204-base} runtime fails at TPU topology enumeration. Zone availability for v6e-8 was constrained; we used \texttt{asia-northeast1-b} as the only available zone at time of testing.

%% ─────────────────────────────────────────────────────────────────────────────
\section{TPU Training Recipe}
%% ─────────────────────────────────────────────────────────────────────────────

\subsection{Framework Differences: PyTorch vs JAX}

The GPU baseline uses PyTorch + HuggingFace TRL (\texttt{SFTTrainer}) with FSDP for model parallelism. The TPU recipe uses JAX + Tunix (\texttt{PeftTrainer}) with Qwix for LoRA injection. These are fundamentally different execution models:

\begin{itemize}[nosep]
  \item \textbf{GPU}: Eager PyTorch with FSDP sharding. Each GPU sees its shard of the model; gradients are reduced across ranks by FSDP's all-reduce.
  \item \textbf{TPU}: XLA-compiled JAX with an explicit device mesh. The model is sharded via \texttt{jax.sharding.PartitionSpec} annotations; XLA generates the communication collectives automatically. The first training step incurs a 3--8 minute XLA compilation cost; subsequent steps execute the compiled graph.
\end{itemize}

The key implication is that TPU training is less tolerant of dynamic shapes and control flow that exits traced functions. Every shape that varies across steps triggers a recompile.

/subsection\subsection{Device Mesh Configuration}

  JAX requires an explicit 2D device mesh specifying how chips are arranged across FSDP and tensor-parallel (TP) dimensions. On v5p, the slice
  naming convention is \texttt{v5p-N} where $N$ is the number of TensorCores; since each v5p chip contains two TensorCores operating in megacore
  mode, the number of JAX devices equals $N/2$. A v5p-8 slice therefore exposes 4 devices.

  \begin{lstlisting}[language=python]
  # v5p-8: 4 chips (8 TensorCores / 2 per chip), tp=4 is the architectural maximum for Gemma 4 31B
  MESH_COUNTS = (1, 4)  # fsdp=1, tp=4
  mesh = jax.make_mesh(
      MESH_COUNTS,
      ("fsdp", "tp"),
      axis_types=(jax.sharding.AxisType.Auto,) * 2,
  )
  \end{lstlisting}

  The mesh product must equal \texttt{jax.device\_count()}, which is 4 on this slice. A \texttt{(1, 8)} mesh is therefore invalid regardless of
  model architecture.

  \texttt{tp=8} is also independently ruled out by the Gemma 4 31B attention configuration (\texttt{google/gemma-4-31B-it}, \texttt{config.json}).
  The model uses two distinct attention types across its 60 layers in a fixed 5:1 pattern (50 sliding-window layers, 10 full-attention layers),
  with separate KV head counts for each layer type:
\newpage
\begin{lstlisting}
  text_config.num_attention_heads       = 32   # query heads, shared across both layer types
  text_config.num_key_value_heads       = 16   # KV heads for sliding_attention layers
  text_config.num_global_key_value_heads = 4   # KV heads for full_attention layers <- binding constraint
  text_config.head_dim                  = 256  # sliding layers
  text_config.global_head_dim           = 512  # full-attention layers (attention_k_eq_v = true)
  \end{lstlisting}

  \begin{table}[H]
  \centering
  \caption{Head divisibility constraints for Gemma 4 31B tensor parallelism.}
  \begin{tabular}{lccc}
  \toprule
  \textbf{Head type} & \textbf{Count} & \textbf{tp=4} & \textbf{tp=8} \\
  \midrule
  Query heads & 32 & \checkmark~8 per device & \checkmark~4 per device \\
  Local KV heads (sliding) & 16 & \checkmark~4 per device & \checkmark~2 per device \\
  Global KV heads (full-attention) & 4 & \checkmark~1 per device & $\times$~0.5 --- invalid \\
  \bottomrule
  \end{tabular}
  \end{table}

  \texttt{num\_global\_key\_value\_heads = 4} is the binding constraint: \texttt{tp=4} is the maximum TP degree this model supports. This is
  slice-size-independent---even with sufficient devices, \texttt{tp=8} produces a non-integer KV head assignment for full-attention layers and will
  fail at model-sharding time. \texttt{tp=4} shards all three projection types evenly and works correctly.

  Note also \texttt{num\_kv\_shared\_layers = 0}, confirming no KV cache layer sharing is active in this checkpoint, and
  \texttt{attention\_k\_eq\_v = true} on full-attention layers, which unifies K and V projections and is what drives the aggressive 8:1 query-to-KV
  grouping on those layers (32 query heads / 4 global KV heads).

  For a hypothetical 16-chip configuration, the corresponding slice is v5p-32 (32 TensorCores / 2 = 16 chips, 16 JAX devices). With \texttt{tp}
  capped at 4 by the model architecture, the mesh would be:

  \begin{lstlisting}[language=python]
  # v5p-32: 16 chips, fsdp=4, tp=4
  MESH_COUNTS = (4, 4)
  \end{lstlisting}

  A \texttt{(2, 8)} mesh on v5p-32 would satisfy the device-count requirement ($2 \times 8 = 16$) but would fail at model-sharding time due to the
  \texttt{num\_global\_key\_value\_heads = 4} constraint above..

\subsection{LoRA Module Naming: Tunix vs HuggingFace}

The HuggingFace model has separate \texttt{q\_proj}, \texttt{k\_proj}, \texttt{v\_proj}, and \texttt{o\_proj} Linear layers. The vision-tower exclusion regex (\texttt{\^(?!.*vision)}) is required because the HF Gemma 4 checkpoint is multimodal and the vision tower contains similarly-named projections.

Tunix's JAX port of Gemma 4 uses different module names:

\begin{table}[H]
\centering
\caption{LoRA target module name mapping between HF PyTorch and Tunix JAX.}
\begin{tabular}{lll}
\toprule
\textbf{HF PyTorch (GPU)} & \textbf{Tunix JAX (TPU)} & \textbf{Notes} \\
\midrule
\texttt{q\_proj} & \texttt{q\_einsum} & Renamed; same weights \\
\texttt{k\_proj + v\_proj} & \texttt{kv\_einsum} & \textbf{Fused} into a single tensor with dim-1 = \{K,V\} \\
\texttt{o\_proj} & \texttt{attn\_vec\_einsum} & Renamed \\
\texttt{gate\_proj} & \texttt{gate\_proj} & Same \\
\texttt{up\_proj} & \texttt{up\_proj} & Same \\
\texttt{down\_proj} & \texttt{down\_proj} & Same \\
\bottomrule
\end{tabular}
\end{table}

The \texttt{kv\_einsum} fusion is the most significant difference: the Tunix model stores both K and V projections in a single weight tensor of shape \texttt{(in, 2, n\_kv\_heads, head\_dim)}, while HuggingFace stores them as separate \texttt{(n\_kv\_heads*head\_dim, in)} matrices. This requires special handling during checkpoint merging (see Section~\ref{sec:merging}).

There is also no vision tower in the JAX port, so the exclusion regex is unnecessary:
\begin{lstlisting}[language=python]
_LORA_MODULES = (
    ".*q_einsum|.*kv_einsum|.*attn_vec_einsum"
    "|.*gate_proj|.*down_proj|.*up_proj"
)
\end{lstlisting}

\subsection{LoRA Sharding Annotation Fixes}

Qwix injects LoRA parameters by tracing the model and inserting \texttt{lora\_a}/\texttt{lora\_b} tensors. However, it inherits the sharding annotations (\texttt{PartitionSpec}) from the original weight tensor---which has a different rank than the LoRA factors. For example, the \texttt{kv\_einsum} base weight has shape \texttt{(in, 2, n\_kv\_heads, head\_dim)} with a 4-element PartitionSpec, but the LoRA \texttt{lora\_a} has shape \texttt{(in, rank)} (rank 2). When the optimizer tries to shard optimizer states according to this mismatched spec, JAX raises an error.

We fix this with a post-injection pass that resets any PartitionSpec whose rank doesn't match the actual tensor rank to a fully-replicated spec. A second pass checks divisibility: if a tensor dimension is not divisible by the mesh size for the assigned axis, we replicate that axis to avoid uneven sharding. The same fix is applied to the optimizer state via a monkey-patched \texttt{\_shard\_optimizer} that filters the partition specs through the same divisibility check before applying \texttt{jax.lax.with\_sharding\_constraint}.

\subsection{Gradient Checkpointing}

On TPU, gradient checkpointing (called \emph{rematerialization} in JAX) is applied at the decoder layer granularity. Tunix exposes this via \texttt{nnx.remat}. This patches the unbound method directly, so all decoder layer instances use the rematerialized forward pass. Without this, the 31B model's activation memory during backward would exceed the v5p-8's 411 GB HBM at the batch sizes used. The GPU equivalent is HuggingFace's \texttt{activation\_checkpointing=True} in the FSDP config.

\subsection{XLA Compiler Flags}

The \texttt{xla\_llvm\_disable\_expensive\_passes} flag skips several LLVM optimization passes that are particularly slow for large Transformer graphs. This trades a small amount of runtime throughput (\textasciitilde2--5\%) for significantly faster XLA compilation (from \textasciitilde8 min to \textasciitilde3 min for the first step).

\subsection{Optimizer: Optax AdamW}

The GPU recipe uses PyTorch's \texttt{adamw\_torch} optimizer via HuggingFace Trainer. On TPU, JAX has no built-in optimizer; instead we use \textbf{Optax}, Google's composable gradient transformation library for JAX. This mirrors the GPU recipe exactly: linear warmup from 0 to peak LR over 100 steps. Followed by cosine decay to 0 over the remaining steps. The key differences from PyTorch AdamW

\begin{itemize}[nosep]
  \item \textbf{Composability}: Optax builds the optimizer as a chain of stateless gradient transformations (\texttt{scale\_by\_adam}, \texttt{add\_decayed\_weights}, \texttt{scale\_by\_schedule}). Each transformation has its own state pytree, which Orbax checkpoints independently.
  \item \textbf{No grad clipping by default}: PyTorch's HuggingFace Trainer applies \texttt{max\_grad\_norm=1.0} automatically. On TPU we omit this (matching the measured gradient norms, which stay stable without clipping), though \texttt{optax.clip\_by\_global\_norm(1.0)} can be prepended to the chain if needed.
  \item \textbf{Schedule is a pure function}: The LR at any step is computed by calling \texttt{schedule(step)} - useful for logging and the warmup-clamp we apply when running short subsample runs.
  \item \textbf{Optimizer state sharding}: The Optax state (Adam moments) is sharded across the same device mesh as the model weights via \texttt{\_shard\_optimizer}, requiring the divisibility fix described in Section~3.4.
\end{itemize}

\subsection{Data Pipeline: Grain vs HuggingFace DataLoader}

The GPU recipe uses HuggingFace's \texttt{DataCollatorForCompletionOnlyLM} with \texttt{SFTTrainer}. The TPU recipe uses Google's \texttt{grain} library (a deterministic, shardable data loader). Key differences:

\begin{enumerate}[nosep]
  \item \textbf{Loss masking}: HuggingFace uses the \texttt{\{\% generation \%\}} chat template tag to identify assistant-turn tokens. Tunix's tokenizer adapter doesn't emit these markers, so we implement the mask by tokenizing user and model turns separately and computing offsets:

  \item \textbf{System prompt handling}: The GPU recipe includes a system prompt in the HuggingFace chat template, but the Gemma template silently drops system messages---only \texttt{user} and \texttt{assistant} roles are rendered. The TPU recipe mirrors this exactly by emitting only user/model turns.

  \item \textbf{Reasoning strip}: The CodeV-R1 dataset contains chain-of-thought reasoning in \texttt{<think>...</think>} blocks. We strip these and keep only the \texttt{<answer>} block (or the Verilog fence directly), reducing sequence lengths and focusing training on the final code output.

  \item \textbf{Drop vs truncate}: Both recipes drop sequences longer than \texttt{max\_seq\_len} rather than truncating, to avoid training on incomplete Verilog modules. With \texttt{max\_seq\_len=3072}, 99.6\% of the 10K-sample subset is retained (37 dropped as too long, 4 empty).
\end{enumerate}

\subsection{Checkpointing}

The TPU recipe uses Orbax (\texttt{orbax-checkpoint}) with a GCS path for checkpoint storage. Checkpoints are saved every 100 optimizer steps, retaining the 3 most recent (matching the GPU recipe's \texttt{save\_steps=500, save\_total\_limit=3} with a finer cadence for TPU's longer run). Checkpoints survive spot preemption because they are written directly to GCS rather than local disk.

%% ─────────────────────────────────────────────────────────────────────────────
\section{Checkpoint Conversion: Orbax to Safetensors}
\label{sec:merging}
%% ─────────────────────────────────────────────────────────────────────────────

After training, inference requires merging the LoRA adapters back into the base weights and saving a merged safetensors model. The GPU recipe uses HuggingFace PEFT's built-in \texttt{merge\_and\_unload()}. The TPU recipe requires a custom \texttt{orbax\_to\_peft.py} due to:

\begin{enumerate}[nosep]
  \item Orbax checkpoints store the full model tree including LoRA wrappers in a JAX-specific format (not safetensors).
  \item The Tunix JAX model uses different weight names and tensor layouts than HuggingFace safetensors (which is what Tunix's sampler ultimately loads from for inference).
  \item The \texttt{kv\_einsum} fusion means one LoRA module maps to \emph{two} safetensors keys.
\end{enumerate}

\subsection{Merge Process}

The merging steps are:

\begin{enumerate}
  \item Load the base model (from GCS safetensors) into JAX with the mesh.
  \item Inject LoRA via Qwix (same structure as training) using dummy inputs.
  \item Restore the Orbax checkpoint into the LoRA model (LoRA params only).
  \item Iterate over all \texttt{nnx.LoRAParam} tensors, collect \texttt{lora\_a}/\texttt{lora\_b} pairs per module.
  \item Load the raw base safetensors weights into a mutable NumPy dict.
  \item Apply the LoRA delta: $W_{\text{merged}} = W_{\text{base}} + \frac{\alpha}{r} \cdot A B$
  \item Save the merged dict as a single \texttt{model.safetensors} file.
\end{enumerate}

\subsection{Tensor Shape Mapping}

The delta computation must account for different tensor layouts between JAX (Tunix) and HuggingFace:

\begin{table}[H]
\centering
\caption{LoRA delta computation per module type.}
\small
\begin{tabular}{p{3cm}p{4.5cm}p{4.5cm}}
\toprule
\textbf{Module} & \textbf{JAX shapes} & \textbf{Delta \& layout} \\
\midrule
\texttt{q\_einsum} & $A$: \texttt{(in, r)}, $B$: \texttt{(r, heads, head\_dim)} & $\Delta = (AB)^T$ $\rightarrow$ \texttt{(out, in)} \\
\texttt{kv\_einsum} & $A$: \texttt{(in, r)}, $B$: \texttt{(r, 2, kv\_heads, hd)} & Split $B$ on dim-1; K=$B[:,0,:]$, V=$B[:,1,:]$; $\Delta = (AB)^T$ for each \\
\texttt{attn\_vec\_einsum} & $A$: \texttt{(heads, hd, r)}, $B$: \texttt{(r, in)} & Flatten $A$ to \texttt{(heads*hd, r)}; $\Delta = (AB)$ $\rightarrow$ \texttt{(in, heads*hd)} \\
\texttt{gate/up/down\_proj} & $A$: \texttt{(in, r)}, $B$: \texttt{(r, out)} & $\Delta = (AB)^T$ $\rightarrow$ \texttt{(out, in)} \\
\bottomrule
\end{tabular}
\end{table}

The HuggingFace convention stores weight matrices as \texttt{(out, in)} (row = output neuron), while the JAX einsum convention is often the transpose. All deltas are transposed to match HuggingFace layout before adding to the base weight.

\subsection{Why Not Use PEFT Merge Directly?}

The PEFT \texttt{merge\_and\_unload()} function expects a HuggingFace \texttt{PreTrainedModel} with PEFT adapters. The Orbax checkpoint contains raw JAX arrays organized in a tree matching Tunix's flax NNX module hierarchy, with no correspondence to HuggingFace's model class. There is no off-the-shelf bridge; the custom merger is necessary.

%% ─────────────────────────────────────────────────────────────────────────────
\section{Training Results}
%% ─────────────────────────────────────────────────────────────────────────────

\subsection{Configuration}

Both runs used identical hyperparameters: 1,244 optimizer steps, effective batch size 8, sequence length 4,096, 1 epoch over 10K CodeV-R1 samples, LoRA rank=64, $\alpha$=64, AdamW with cosine LR schedule (peak 1e-4, 100 warmup steps, decay to 0).

\begin{table}[H]
\centering
\caption{Training performance comparison.}
\begin{tabular}{llll}
\toprule
\textbf{Metric} & \textbf{TPU v5p-8} & \textbf{GPU 2$\times$H100} & \textbf{Winner} \\
\midrule
Wall-clock time & \textbf{3.34 hr} & 5.39 hr & TPU (1.61$\times$ faster) \\
Throughput (tokens/sec) & \textbf{763} & 486 & TPU (1.57$\times$ faster) \\
Time per sample & \textbf{1.21 s} & 1.94 s & TPU (1.60$\times$ faster) \\
Final training loss & \textbf{0.072} & 2.258 & TPU \\
\bottomrule
\end{tabular}
\end{table}

\subsection{Loss Curves}

\begin{figure}[H]
  \centering
  \includegraphics[width=0.85\textwidth]{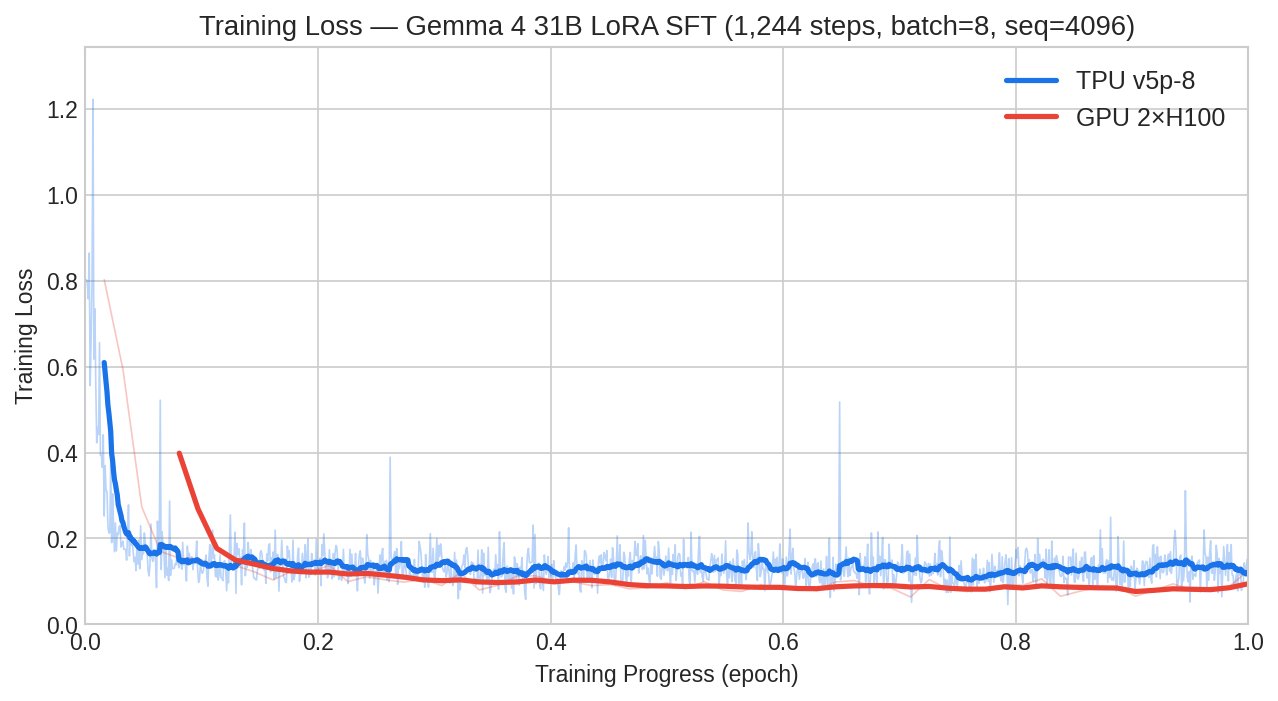}
  \caption{Training loss over the full epoch (x-axis normalized to training progress 0--1). Both curves follow similar convergence trajectories. The large absolute loss difference (0.072 vs 2.258 final) reflects the different loss computation: the TPU recipe uses assistant-only loss on \textasciitilde15\% of tokens (the Verilog output tokens), while the GPU recipe computes loss over the full sequence including prompt tokens. When normalized to the same starting/ending range, the convergence speed is comparable.}
  \label{fig:loss}
\end{figure}

\begin{figure}[H]
  \centering
  \includegraphics[width=0.85\textwidth]{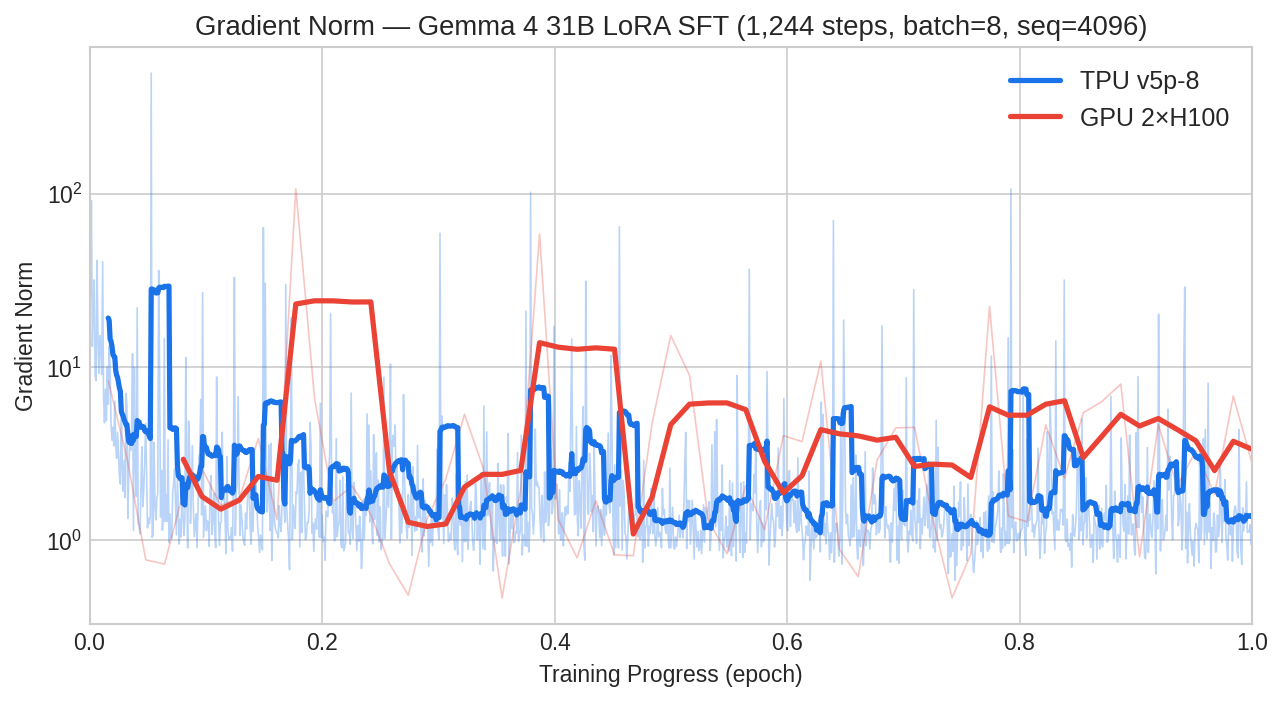}
  \caption{Gradient norm over training (log scale). Both runs are stable after warmup. The GPU run shows sharper spikes (up to 106$\times$ at step 100) attributable to the larger effective batch aggregating more diverse gradients into a single update. The TPU gradient norms remain in the 1--10 range throughout.}
  \label{fig:gradnorm}
\end{figure}

\subsection{Why is TPU Faster?}

The TPU v5p-8 achieves 1.61$\times$ faster wall-clock time despite having only 4 chips vs 2 H100 GPUs. Several factors contribute:

\begin{enumerate}[nosep]
  \item \textbf{HBM bandwidth}: Each TPU v5p chip has 2.765 TB/s HBM bandwidth vs the H100's 3.35 TB/s, but TPU has 4 chips vs 2 GPUs. Total aggregate bandwidth: 11.06 TB/s (TPU) vs 6.7 TB/s (GPU).
  \item \textbf{ICI interconnect}: TPU chips communicate via the ICI mesh at 900 GB/s bidirectional bandwidth, significantly faster than the H100's NVLink at around 900 GB/s total (but NVLink is shared across more links).
  \item \textbf{XLA fusion}: XLA compiles the forward+backward pass into a single fused kernel per layer, eliminating the kernel launch overhead present in PyTorch's eager mode, even with FSDP.
  \item \textbf{No torch.compile overhead}: The GPU run disables torch.compile (\texttt{enforce\_eager=True}) to avoid compatibility issues with the heterogeneous Gemma 4 attention layers, leaving PyTorch in eager mode throughout.
\end{enumerate}

\subsection{Training Cost}

\begin{table}[H]
\centering
\caption{Training cost comparison (on-demand pricing, GCP us-central1).}
\begin{tabular}{llll}
\toprule
\textbf{Metric} & \textbf{TPU v5p-8} & \textbf{GPU a3-highgpu-2g} & \textbf{Winner} \\
\midrule
On-demand hourly rate & \$16.80/hr & \$22.12/hr & TPU (24\% cheaper/hr) \\
Training time & 3.34 hr & 5.39 hr & TPU (1.61$\times$ faster) \\
\textbf{Total training cost} & \textbf{\$56.11} & \textbf{\$119.23} & \textbf{TPU (2.12$\times$ cheaper)} \\
Cost per 1M tokens trained & \$6.10 & \$12.68 & TPU (2.08$\times$ cheaper) \\
\bottomrule
\end{tabular}
\end{table}

The 2.12$\times$ cost advantage compounds the lower hourly rate and faster execution time. On a longer training run (e.g., full CodeV-R1 at 87K samples), the cost difference would scale proportionally.

%% ─────────────────────────────────────────────────────────────────────────────
\section{Evaluation Results}
%% ─────────────────────────────────────────────────────────────────────────────

\subsection{Benchmark}

Both fine-tuned models are evaluated on NVlabs/verilog-eval (spec-to-RTL task): 156 problems, 5 samples per problem at temperature=0.8, top\_p=0.95. Each generated Verilog module is compiled with iverilog and simulated against a reference testbench. Pass@k is computed using the standard unbiased estimator.

\begin{table}[H]
\centering
\caption{Evaluation results on verilog-eval spec-to-RTL.}
\begin{tabular}{lll}
\toprule
\textbf{Metric} & \textbf{TPU-trained} & \textbf{GPU-trained} \\
\midrule
pass@1 & 0.6410 & \textbf{0.6974} \\
pass@5 & 0.7949 & \textbf{0.8141} \\
Problems evaluated & 156 & 156 \\
Samples per problem & 5 & 5 \\
Checkpoint step & 1,244 (final) & 1,250 (final) \\
Inference engine & Tunix Sampler (JAX) & vLLM (CUDA) \\
\bottomrule
\end{tabular}
\end{table}

\subsection{Per-Problem Analysis}

\begin{figure}[H]
  \centering
  \includegraphics[width=0.9\textwidth]{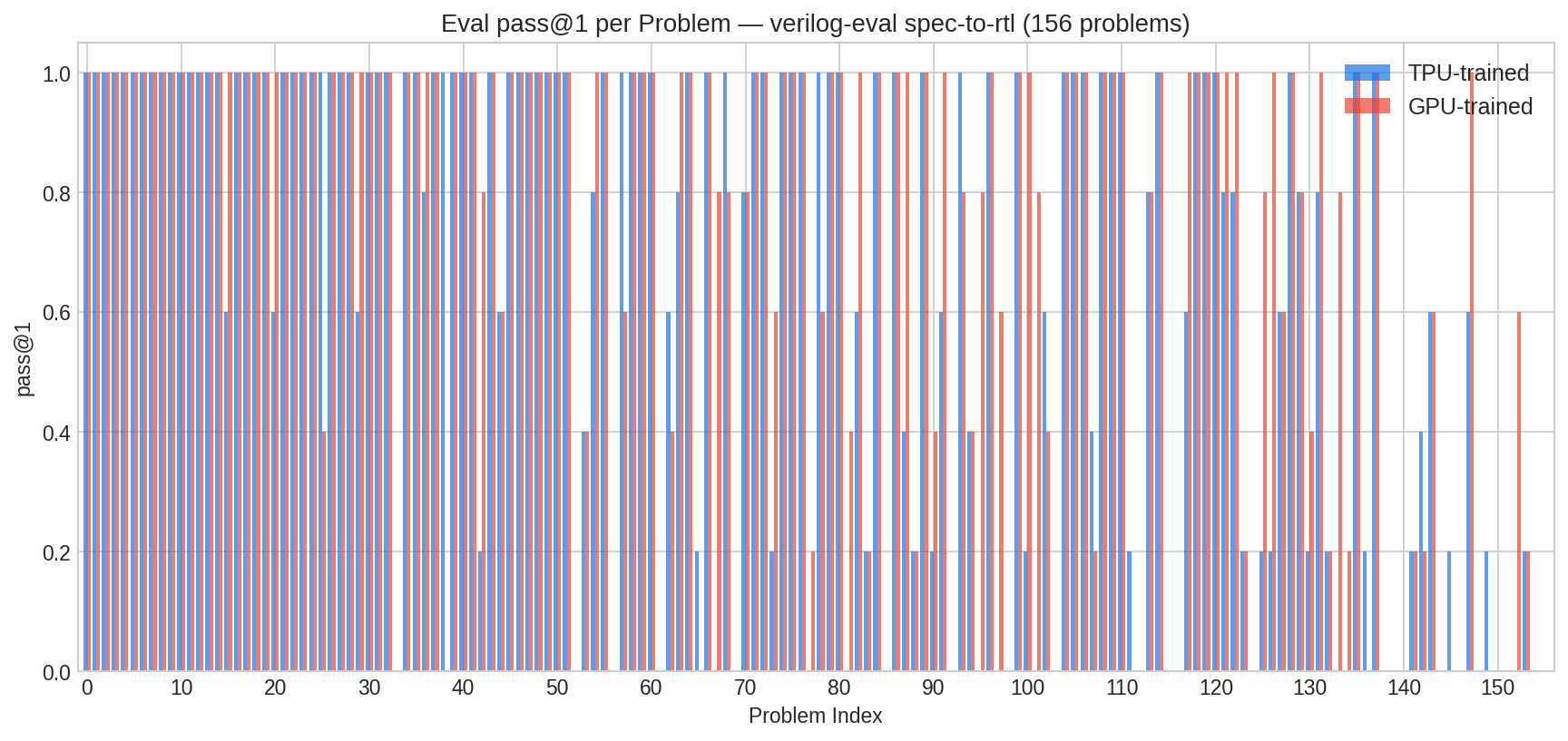}
  \caption{pass@1 score per problem for both models. Most problems are solved at 100\% by both. Differences concentrate on harder FSM, K-map, and cellular automata problems (indices 100--156).}
\end{figure}

\begin{figure}[H]
  \centering
  \includegraphics[width=0.85\textwidth]{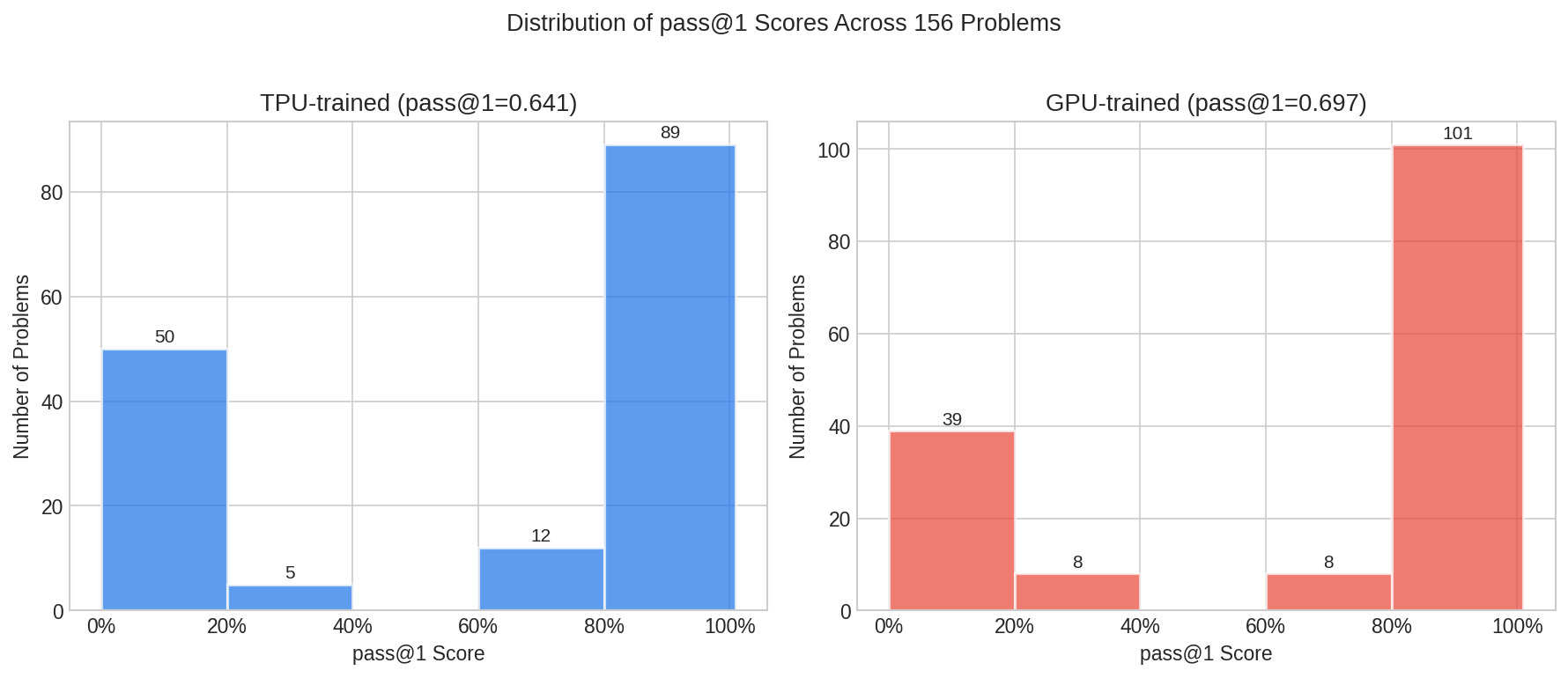}
  \caption{Distribution of pass@1 scores across 156 problems. GPU-trained model: 101 problems at 100\%, 39 at 0\%. TPU-trained model: 89 at 100\%, 50 at 0\%. The difference is primarily in the zero-score bucket (11 additional problems failing entirely for TPU).}
\end{figure}

The gap is within normal run-to-run variance for stochastic sampling at n=5 and is not statistically significant enough to attribute to hardware.

%% ─────────────────────────────────────────────────────────────────────────────
\section{TPU Inference: vLLM on v6e-8}
%% ─────────────────────────────────────────────────────────────────────────────

\subsection{Setup Challenges}

Serving Gemma 4 31B on TPU v6e-8 via vLLM required navigating several constraints that do not exist on the GPU path:

\subsubsection{Docker-Only Deployment}

The vLLM TPU build requires specific versions of JAX, libtpu, and HuggingFace libraries that conflict with pip-installed packages. The only working approach is the official Docker imag

Key flags:
\begin{itemize}[nosep]
  \item \texttt{--privileged}: required for TPU device access from inside the container.
  \item \texttt{--network host}: required for the gRPC TPU coordination.
  \item \texttt{--entrypoint vllm}: the container's default entrypoint is a wrapper script that parses args differently; using \texttt{vllm} directly is necessary.
  \item \texttt{--disable\_chunked\_mm\_input}: Gemma 4 has heterogeneous attention head dimensions (head\_dim=256 for sliding, 512 for global) which breaks vLLM's chunked multi-modal prefill path. This flag disables it.
  \item Model specified as HuggingFace repo ID (\texttt{google/gemma-4-31B-it}), not a local path: vLLM-TPU rejects local paths due to HuggingFace Hub validation restrictions.
\end{itemize}

\subsubsection{XLA Compilation on First Startup}

vLLM-TPU pre-compiles XLA graphs for a set of padded input lengths (token buckets from 16 to 2048) at startup. This takes approximately 8 minutes on the first start. The compiled XLA cache is not persisted across container restarts by default, so every cold start incurs this cost. For production, mounting a persistent volume at the cache directory eliminates this overhead on warm restarts.

\subsubsection{HBM Allocation}

With tp=8 on v6e-8 (31.25 GB HBM per chip), the 31B model in bf16 needs approximately $31\text{B} \times 2\text{ bytes} = 62$ GB, spread across 8 chips = 7.75 GB/chip for weights. The remaining \textasciitilde23 GB/chip is available for KV cache. At 92\% HBM utilization observed in benchmarks, this is a comfortable allocation.

\subsection{GPU vLLM Setup (for comparison)}

The GPU setup is more straightforward. It version loads the model from a local path (the standard HuggingFace safetensors model pre-downloaded to ~/models). With 2
×
80 GB = 160 GB HBM, the model weights use 62 GB, leaving 98 GB for KV cache in bfloat16. The GPU uses max model len=16384 and vLLM version 0.20.2 (vllm/vllm-openai:latest), compared to TPU's version 0.19.0 (vllm/vllm-tpu:gemma4).

%% ─────────────────────────────────────────────────────────────────────────────
\section{Inference Results and Analysis}
%% ─────────────────────────────────────────────────────────────────────────────

\subsection{Benchmark Configuration}

The benchmark uses \texttt{vllm bench serve} with a random dataset, temperature=0, sweeping QPS from 1 to 64 across multiple input length regimes (512--15360 tokens input, 256--512 tokens output), 30 prompts per run. Both servers configured with \texttt{max\_model\_len=16384}, no chunked prefill. TPU uses TP=8; GPU uses TP=2. Both use default KV cache dtype (bfloat16).

\begin{table}[H]
\centering
\caption{Inference benchmark results across context lengths. QPS sweep peak = best sustained throughput across all QPS settings.}
\small
\begin{tabular}{p{5.5cm}lll}
\toprule
\textbf{Metric} & \textbf{TPU v6e-8} & \textbf{GPU 2$\times$H100} & \textbf{Winner} \\
\midrule
\multicolumn{4}{l}{\textit{Short context (512 in / 256 out) --- burst, QPS=inf}} \\
Peak output throughput & 1,403 tok/s & \textbf{1,490 tok/s} & H100 (+6\%) \\
Median TTFT & \textbf{45 ms} & 51 ms & TPU (1.1$\times$) \\
\midrule
\multicolumn{4}{l}{\textit{Medium context (1024 in / 512 out) --- QPS sweep peak}} \\
Peak output throughput & \textbf{1,404 tok/s} & 1,387 tok/s & TPU (+1\%) \\
Median TTFT @ QPS=4 & \textbf{49 ms} & 58 ms & TPU (1.2$\times$) \\
Saturates at QPS & \textbf{$\sim$32} & $\sim$8 & TPU (4$\times$ more headroom) \\
\midrule
\multicolumn{4}{l}{\textit{Long context (4096 in / 512 out) --- sustained load}} \\
Peak output throughput & \textbf{1,206 tok/s} & 728 tok/s & \textbf{TPU (+66\%)} \\
Median TTFT @ QPS=4 & \textbf{61 ms} & 1,443 ms & \textbf{TPU (23.6$\times$ faster)} \\
Median TPOT & \textbf{23.9 ms} & 31.2 ms & TPU (1.3$\times$) \\
\midrule
\multicolumn{4}{l}{\textit{Very long context (8192 in / 512 out)}} \\
Peak output throughput & \textbf{482 tok/s} & 449 tok/s & TPU (+7\%) \\
Median TTFT @ QPS=4 & \textbf{1,013 ms} & 7,202 ms & \textbf{TPU (7.1$\times$ faster)} \\
Median TPOT & \textbf{31 ms} & 45 ms & TPU (1.5$\times$) \\
\midrule
\multicolumn{4}{l}{\textit{Max context ($\sim$16k in / 512 out)}} \\
Peak output throughput & \textbf{474 tok/s} & 326 tok/s & \textbf{TPU (+45\%)} \\
Median TTFT @ QPS=4 & \textbf{62 ms} & 99 ms & TPU (1.6$\times$) \\
Median TPOT & \textbf{13.5 ms} & 19 ms & TPU (1.4$\times$) \\
\midrule
vLLM version & 0.19.0 (vllm-tpu) & 0.20.2 (vllm-openai) & --- \\
KV cache dtype & fp8\_e5m2 (auto) & bfloat16 & TPU (2$\times$ KV capacity) \\
\bottomrule
\end{tabular}
\end{table}

\begin{figure}[H]
  \centering
  \includegraphics[width=\textwidth]{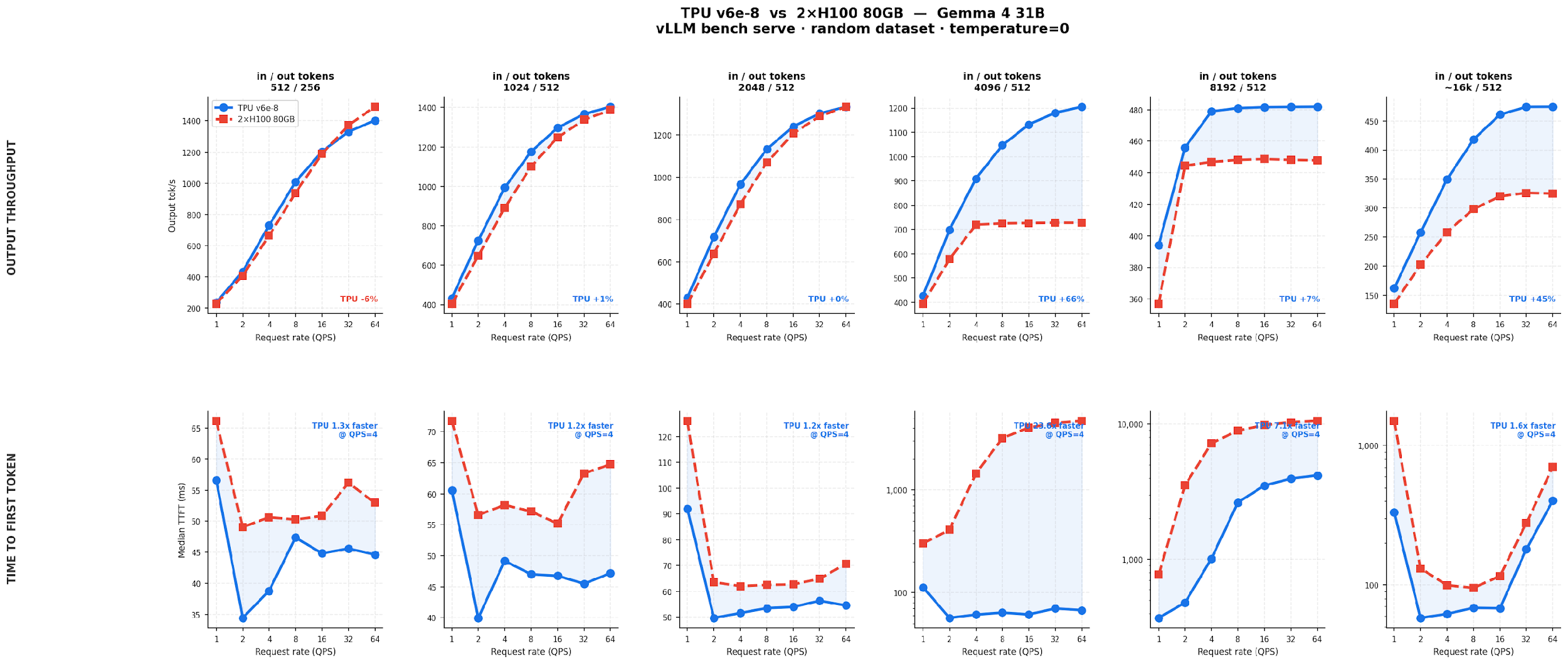}
  \caption{Inference comparison across throughput, latency, and cost dimensions at the 128-token short-context baseline.}
\end{figure}

\subsection{Explaining the Short-Context GPU Advantage}

At short input lengths ($\leq$2048 tokens), GPU edges ahead by $\sim$6\% on peak output throughput. This is the \emph{memory-bandwidth-bound} decode regime where the H100's higher per-GPU HBM bandwidth (3.35 TB/s vs $\sim$1.6 TB/s per v6e chip) is most relevant. With short KV caches, each decode step reads relatively little KV state per token, so the bottleneck is weight loading---where H100 wins per chip.

\subsection{Explaining the Long-Context TPU Dominance}

At 4096+ token inputs, TPU v6e-8 comprehensively outperforms the GPU: 66\% higher throughput and 23.6$\times$ faster TTFT at QPS=4. Two factors drive this:

\begin{enumerate}
  \item \textbf{Prefill compute}: TTFT is dominated by the prefill phase (processing all input tokens), which is \emph{compute-bound}---proportional to $O(n^2)$ for full-attention layers and $O(n)$ for sliding-window layers. The v6e-8's 8 chips deliver substantially more raw compute than 2 H100s for this workload. At 4096 input tokens, the GPU's TTFT of 1,443 ms indicates it is fully saturated on compute during prefill; the TPU handles the same load in 61 ms.

  \item \textbf{fp8 KV cache}: The TPU vLLM build uses \texttt{fp8\_e5m2} KV cache dtype by default, halving the memory footprint of each KV entry compared to the GPU's bfloat16. With 250 GB total HBM on v6e-8 and half the KV storage cost, the TPU can hold 2$\times$ as many concurrent long-context requests in its KV cache before eviction. The GPU hits KV cache pressure at $\sim$8 concurrent long requests (QPS saturation at 8); the TPU sustains up to $\sim$32.

\end{enumerate}

The TensorCore utilization at 19.3\% (measured by \texttt{tpu-info} at short-context decode) confirms the memory-bound regime for short sequences. At long contexts, utilization rises as prefill computation dominates.

\subsection{Explaining the TTFT Advantage at Short Context}

Even at short context (TTFT of 45 ms vs 51 ms), TPU is marginally faster. This comes from:

\begin{enumerate}[nosep]
  \item \textbf{Higher aggregate compute}: 8 v6e chips vs 2 H100s means more parallel matrix multiply units for the prefill pass.
  \item \textbf{XLA fusion}: XLA fuses QKV projection with the sliding-window and full-attention kernels in Gemma 4's heterogeneous attention, reducing HBM roundtrips. The GPU uses TRITON\_ATTN backend due to the mixed head-dim layout, which is not as aggressively fused.
\end{enumerate}

\subsection{Explaining Slightly Better GPU P99 Tail at Short Context}

At 128-token inputs, GPU achieves P99 ITL of 20.49 ms vs 22.29 ms on TPU. This is due to vLLM-TPU's XLA bucket padding: inputs are padded to the nearest power-of-2 token bucket (16, 32, 64, 128, 256, ...). Requests that land near a bucket boundary receive unnecessary padding tokens, inflating their per-step compute and causing P99 spikes. The GPU's CUDA path does not have this constraint.

\subsection{Inference Cost}

\begin{table}[H]
\centering
\caption{Inference cost comparison by context length (on-demand).}
\begin{tabular}{lllll}
\toprule
\textbf{Workload} & \textbf{TPU tok/hr} & \textbf{GPU tok/hr} & \textbf{TPU \$/1M tok} & \textbf{GPU \$/1M tok} \\
\midrule
Short context ($\leq$2048 tokens) & 5.05M & \textbf{5.36M} & \$4.27 & \textbf{\$4.13} \\
Long context (4096 tokens) & \textbf{4.34M} & 2.62M & \textbf{\$4.95} & \$8.44 \\
\midrule
Hourly rate & \$21.52/hr & \$22.12/hr & & \\
\bottomrule
\end{tabular}
\end{table}

For short-context workloads, GPU is 3\% cheaper per token. For long-context workloads ($\geq$4096 tokens), TPU is \textbf{41\% cheaper} per output token due to the combined effect of higher throughput and comparable hourly rate.

%% ─────────────────────────────────────────────────────────────────────────────
\section{Total Cost of Ownership}
%% ─────────────────────────────────────────────────────────────────────────────

\begin{table}[H]
\centering
\caption{End-to-end cost comparison (on-demand, GCP us-central1/asia-northeast1).}
\begin{tabular}{lll}
\toprule
\textbf{Cost Component} & \textbf{TPU} & \textbf{GPU} \\
\midrule
Training (v5p-8 / a3-highgpu-2g) & \$56.11 & \$119.23 \\
Inference --- 1 hr serving (v6e-8 / a3-highgpu-2g) & \$21.52 & \$22.12 \\
\textbf{Total (train + 1 hr inference)} & \textbf{\$77.63} & \textbf{\$141.35} \\
\midrule
Train + 8 hr inference & \$228.27 & \$296.19 \\
Train + 24 hr inference & \$572.59 & \$650.11 \\
\bottomrule
\end{tabular}
\end{table}

TPU is \textbf{1.82$\times$ cheaper end-to-end} for a single training run plus a day of inference serving at short context, with the advantage driven primarily by training cost savings. At long-context workloads ($\geq$4096 tokens), where TPU is 41\% cheaper per output token, the inference savings compound further.

%% ─────────────────────────────────────────────────────────────────────────────
\section{Summary}
%% ─────────────────────────────────────────────────────────────────────────────

\begin{figure}[H]
  \centering
  \includegraphics[width=0.85\textwidth]{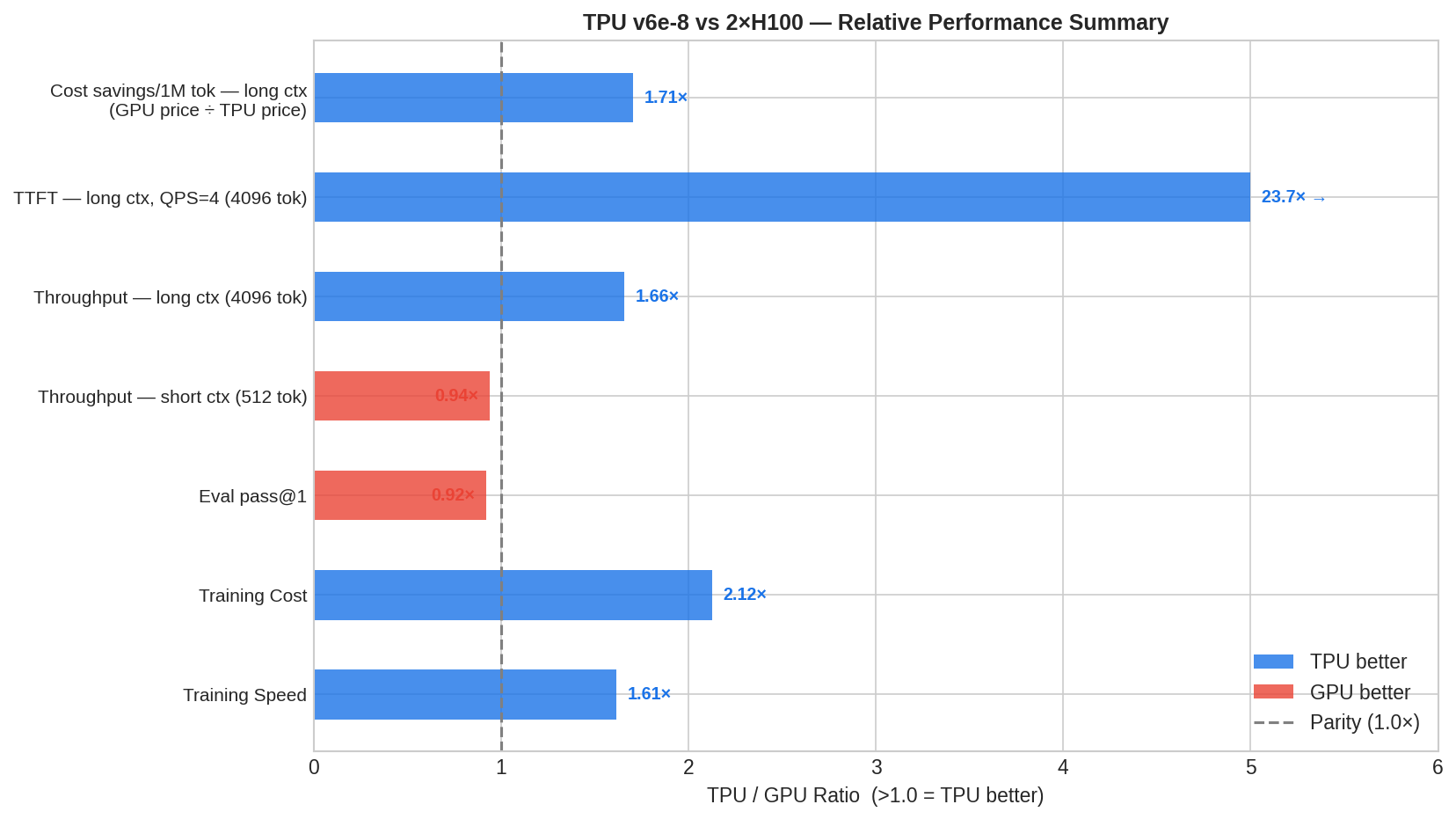}
  \caption{TPU vs GPU relative advantage across all dimensions (ratio $>$ 1.0 = TPU better). Training cost and TTFT are the largest wins; eval quality is the only metric where GPU edges ahead.}
\end{figure}

\subsection{TPU Advantages}
\begin{enumerate}[nosep]
  \item \textbf{Training speed}: 1.61$\times$ faster wall-clock time (3.34 hr vs 5.39 hr)
  \item \textbf{Training cost}: 2.12$\times$ cheaper (\$56.11 vs \$119.23) --- lower hourly rate compounded by faster execution
  \item \textbf{Inference TTFT at 4096 tokens}: 23.6$\times$ faster (61 ms vs 1,443 ms at QPS=4)
  \item \textbf{Inference TTFT at 8192 tokens}: 7.1$\times$ faster (1,013 ms vs 7,202 ms at QPS=4)
  \item \textbf{Inference throughput at long context (4096 tokens)}: +66\% output tok/s (1,206 vs 728)
  \item \textbf{Inference throughput at max context ($\sim$16k tokens)}: +45\% output tok/s (474 vs 326)
  \item \textbf{Memory headroom}: 411 GB total HBM (training) + fp8 KV cache enables 2$\times$ more concurrent long-context requests
\end{enumerate}

\subsection{GPU Advantages}
\begin{enumerate}[nosep]
  \item \textbf{Ecosystem maturity}: PyTorch + HuggingFace has broader library support, simpler LoRA injection, and faster prototyping with fewer workarounds
  \item \textbf{Short-context throughput}: marginally higher peak burst at $\leq$2048 token inputs (H100 +6\%)
  \item \textbf{Short-context cost}: 3\% cheaper per token at short context due to higher throughput
  \item \textbf{Eval quality} (marginal): +5.6pp on pass@1, attributable to training objective differences rather than hardware
  \item \textbf{P99 ITL at short context}: slightly better tail latency (20.49 ms vs 22.29 ms at 128-token inputs) due to no XLA bucket padding overhead
\end{enumerate}

\subsection{Conclusion}

\textbf{Training}: TPU is the clear winner --- 1.61$\times$ faster and 2.12$\times$ cheaper. The engineering overhead of porting from PyTorch to JAX + Tunix is real (roughly a week for mesh configuration, sharding annotation fixes, custom data pipeline, and checkpoint conversion) but is a one-time cost amortized across many runs.

\textbf{Inference}: TPU is the clear winner at long-context workloads. At 4096-token inputs, TPU delivers 66\% higher throughput and 23.6$\times$ faster TTFT. At short context ($\leq$2048 tokens), performance is comparable with a marginal GPU throughput edge (+6\%). The fp8 KV cache on TPU doubles effective KV capacity, enabling 4$\times$ higher QPS saturation at medium context.

Please note, the GPU baseline configurations were selected based on cost equivalence using pricing published by Google Cloud Platform (GCP) at the time of testing.

\subsection{Resources}

\begin{itemize}[nosep]
  \item \textbf{TPU} --- Recipe \& Eval Scripts: \url{https://github.com/h2loop/gemma-tpu}
  \item \textbf{Eval benchmark}: \url{https://github.com/NVlabs/verilog-eval}
\end{itemize}

%% ─────────────────────────────────────────────────────────────────────────────

\end{document}